\documentclass[twocolumn, 
prd, nofootinbib, 
showpacs, 
preprintnumbers, amsmath, amssymb]
              {revtex4}

\usepackage[sort&compress]{natbib}
\usepackage{graphicx}

\begin{document}

\title{A gauge model for right handed neutrinos as dark matter }

\date{February 2012}

\author{ R. J. Hern\'andez-Pinto}
\email{rhernand@fis.cinvestav.mx} \affiliation{ Departamento de 
F\'isica, Centro de Investigaci\'on y de Estudios Avanzados del I.P.N., 
Apdo. Post. 14-740, 07000 M\'exico D.F., M\'exico. }

\author{ A. P\'erez-Lorenzana}
\email{aplorenz@fis.cinvestav.mx} \affiliation{ Departamento de 
F\'isica, Centro de Investigaci\'on y de Estudios Avanzados del I.P.N., 
Apdo. Post. 14-740, 07000 M\'exico D.F., M\'exico. }

\begin{abstract}
$B-L$ is known to be a symmetry somewhat linked to the origin of neutrino masses, which turns out to be anomaly free upon the sole introduction of right handed neutrinos. We suggest a simple extension of the electroweak group, $SU(2)_L\times U(1)_Y\times U(1)_{B-L}$, where the breaking  of $U(1)_{B-L}$ symmetry  provides masses for right handed neutrinos at an acceptable range for them to be Dark Matter candidates. We review the cosmological constraints for this type of models, in order to find the restrictions on the parameters of the model. We study the contributions due to the $B-L$ current in the process $e^+e^- \rightarrow e^+e^-$ at low energies.
\end{abstract}

\maketitle

\section{Introduction}

There is compelling evidence that the Standard Model (SM) of strong and
electroweak interactions is not complete. There are several experimental and
observational facts that cannot be explained within the SM. These are neutrino
oscillations~\cite{nuosc:2008}, the presence 
of Dark Matter (DM) in the Universe~\cite{dm}, the baryon asymmetry of the Universe, its flatness, and the existence of cosmological perturbations necessary for structure formation~\cite{lcdm}.
Neutrino oscillations give indirect proof that neutrinos are massive. Their mass, however, needs that the SM be extended at least to incorporate right handed neutrinos (RHN), and the simplest way to understand the tiny scales required for neutrino oscillations is to implement the see-saw mechanism~\cite{seesaw, Valle}. The simplest extension to the SM that includes the seesaw Lagrangian structure is written as 
\begin{eqnarray}\label{1}
\mathcal{L}_{\nu MSM}&=&\mathcal{L}_{SM}+\bar{N}_I i \partial_{\mu}
\gamma^{\mu}N_I\nonumber \\
&&-\mathcal{K}_{\alpha I}\bar{L}_{\alpha}N_I\tilde{\Phi}-\frac{M_I}{2}\bar{N}^c_I N_I + h.c., 
\end{eqnarray}
where $\mathcal{L}_{SM}$ is the Lagrangian of the Standard Model,
$\tilde{\Phi}_i=\epsilon_{ij}\Phi^*_j$ and $L_{\alpha}$, $(\alpha = e, \mu, \tau )$ are the Higgs and lepton doublets, respectively, and both Dirac $(M^D = \mathcal{K} \langle \Phi \rangle)$ and Majorana $(M_I )$ masses for neutrinos are introduced, where $I=1,2,3$ stands for the number of RHN species. The theory based on this Lagrangian
has been called the  $\nu$-Minimal Standard Model ($\nu$MSM)~\cite{Sha205}. In comparison with the SM, the $\nu$MSM contains 
18 new parameters and interestingly enough, it appears to have the
potential to provide some explanations to the above mentioned problems~\cite{Sha205}. 

The new parameters of the $\nu$MSM can describe any pattern of masses and
mixings of active neutrinos, which is characterized by 9 parameters only.
Despite of this freedom, the absolute scale of active neutrino masses can be
pin pointed in the $\nu$MSM from cosmology and astrophysics of dark matter
particles~\cite{Sha205, Boy106, Boy206, Lai06, Lai07}: one of the active
neutrinos must have a mass smaller than $\mathcal{O}(10^{-5} )$ eV. The choice
of the small mass scale for singlet fermions leads to small values of the Yukawa
coupling constants,on the level of $10^{-6} - 10^{-12}$, which is crucial for
the explanation of dark matter and baryon asymmetry of the Universe.

Although the $\nu$MSM does not have any extra stable particle in comparison with the SM, the lightest singlet fermion, $N_1$ , may have a life-time greatly exceeding the age of the Universe and thus play a role of a dark matter particle~\cite{Dod94, Shi99, Dol02, Aba01,Khlopov}. 
DM sterile neutrinos can be produced in the early Universe via active-sterile neutrino transition~\cite{Dod94}; via resonant active-sterile neutrino oscillations in the presence of lepton asymmetries~\cite{Shi99}; or during inflation. DM sterile neutrino may also have other interesting cosmological applications~\cite{Kus97}, in particular for the understanding of baryon asymmetry of the Universe. On the other hand, the baryon ($B$) and lepton ($L$) numbers are not conserved in the $\nu$MSM. The lepton number is violated by the Majorana neutrino masses, while $B + L$ is broken by its anomaly. As a result, the sphaleron processes with baryon number non-conservation~\cite{Kuz85} are in thermal equilibrium for temperatures 100 GeV $<T<$ 1012 GeV. As for CP-breaking, the $\nu$MSM contains 6 CP-violating phases in the lepton sector and a Kobayashi-Maskawa phase in the quark sector. This makes two of the Sakharov conditions~\cite{Sak67} for baryogenesis satisfied. Similarly to the SM, this theory does not have an electroweak phase transition with allowed values for the Higgs mass~\cite{Kaj96}, making impossible the electroweak baryogenesis, associated with the non-equilibrium bubble expansion. However, the $\nu$MSM contains extra degrees of freedom -sterile neutrinos- which may be out of thermal equilibrium exactly because their Yukawa couplings to ordinary fermions are very small. The latter fact is a key point for the baryogenesis in the $\nu$MSM~\cite{Akh98, Asa05}, ensuring the validity of the third Sakharov condition. In Ref.~\cite{Asa05} it was shown that the $\nu$MSM can provide simultaneous solution to the problem of neutrino oscillations, dark matter and baryon asymmetry of the Universe. 


However, this theory breaks explicitly the global symmetry $B-L$, due to the term $\bar{N}^c_IN_I$, which also represents the introduction by hand of a fermion mass in the scheme, against the established lore  in the standard theory, which claims that fermion as well as gauge boson masses are due to the Higgs mechanism. We consider this as a serious theoretical draw back of the model. This is, of course, a known fact that usually drives theorist to consider the left-right extensions to the standard model (see for instance Ref.~\cite{Moh81}). Nevertheless, which seems quite appealing to exploit is the fact that solely introducing three families of RHN's in the SM particle content comes with the extra free ingredient of making $B-L$ an anomaly free symmetry, without any further requirement. Interestingly enough, keeping the physical connection between the mass of RHN's and the Higgs Mechanism, suggests the simplest gauge group $SU(2)_L\times U(1)_Y\times U(1)_{B-L}$ as a natural symmetry which extends the electroweak group of SM, and which should contains most of the features of the $\nu$MSM. In here, $B-L$ turns out to be unrelated to hypercharge, and thus, the model offers a completely  different route to understand the origin of see-saw terms, and, of course, the gauge origin of the $\nu$MSM, than that provided by left-right type models. To explore some of the features of such gauge extension is the main goal of the present work.
 
This paper is organized as follows, first we  present the model $SU(2)_L\times
U(1)_Y\times U(1)_{B-L}$ and  give the justifications for adding the
$U(1)_{B-L}$ gauge symmetry to the electroweak group. Then, as the new
interactions introduced on the model for RHN's do require us to analyze their
potential over the former $\nu$MSM considerations for DM,  we check the
cosmological constraints on the model. And finally, we discuss the
considerations under which the model could be consistent to collider results. In
particular we focus on low energy Bhabha scattering, which could constrain a
weakly interacting light $B-L$ boson, imposing preliminary bounds on the gauge
coupling to mass ratio. 


\section{The $B-L$ model}
The model under consideration is based in the gauge symmetry group $SU(2)_L\times U(1)_Y\times U(1)_{B-L}$, and, as we have already stated, it is motivated on the idea that the Higgs mechanism should be regarded as the fundamental mechanism for the generation of masses. Matter content is the same as for the SM, but for the addition of a RHN, $N$, per family, and a SM singlet scalar, $\sigma$, that is going to be used to implement spontaneous symmetry
breaking. So, this model contains the following particles,
\begin{eqnarray}
&L& \sim (\textbf{2},-1,-1) ;\nonumber\\
&e_R&\sim (\textbf{1},-2,-1);\nonumber\\
&N&\sim (\textbf{1},0,-1);\\
&H&\sim (\textbf{2},1,0) ;\nonumber\\
&\sigma&\sim (\textbf{1},0,2) ;\nonumber
\end{eqnarray}
where the numbers label the Weyl representation of $(SU(2)_L, U(1)_Y,U(1)_{B-L})$, and family indices are understood. It is not difficult to check that, with this fermion content, $B-L$ is anomaly free~\cite{deq}, and that Yukawa couplings, 
\begin{eqnarray}\label{Yukawa}
 y_I\sigma \bar{N}^c_I N_I,
\end{eqnarray} 
will generate the RHN mass terms of  the $\nu$MSM, 
$M_I/2=y_I\langle\sigma\rangle$, upon spontaneous symmetry breaking, thus,
relating the RHN mass scale to the physical scale at which $B-L$ is being
broken. It is important to notice that $B-L$ in this model appears as a symmetry
that is orthogonal to the hypercharge, and therefore with no relation,
whatsoever, to the electric charge. This is a distinctive feature that makes the
study of the model worthy. 

Several models exist on the literature that had added to the SM, a $U(1)$ gauge
symmetry~\cite{Pat74, Moh75, Moh175, Sen75, Car94,App03,Dob05}; those models
assume that the range of the breaking scale of the symmetry is bigger than few
hundred GeV~\cite{Mar83, Jen87}, which is usually due to the  mixings among the
associated $B-L$ gauge bosons and the standard $Z$ boson, for which the usual
bounds on exotic $Z'$ searches apply~\cite{zp}. Without the tree level mixing,
it is clear those bounds do not apply straightforwardly. In contrast, the 
search of $B-L$ gauge boson, $Z_{B-L}$, through dilepton, or dijet production in
colliders, seems more promising. A complete analysis for extra $Z'$ gauge bosons
from a collider point of view can be found in Ref.~\cite{Carena}. In particular, for
a vector-like $U(1)_{B-xL}$ gauge boson coupled to left and right
handed electrons, one can extract the lower bound due to LEP experiments,
\begin{eqnarray}
M_{Z'}\geq |x|g_{Z'} \times (6 \mbox{ TeV}).
\label{zpbound}
\end{eqnarray}
Thus, in a $B-L$ model there is still a wide range allowed for the mass of the extra gauge boson depending on the coupling. So, we can stretch the LEP limit on the ignorance of the value of gauge coupling associated to $U(1)_{B-L}$. Therefore, the possibility that this gauge group might be broken at scales similar to electroweak scale is yet present, and worth of consideration.

In the model described here, it is supposed the existence of two Higgs fields, one which breaks the electroweak sector ($SU(2)_L\times U(1)_Y$), $H(x)$, and the second which breaks the $U(1)_{B-L}$ symmetry, $\sigma (x)$. The most general lagrangian which should be used to implement symmetry breaking is 
\begin{eqnarray}
\mathcal{L}&=&(\partial_{\mu}H)^{\dagger}(\partial^{\mu}H)-m_H^2|H|^2+\frac{\alpha}{4}|H|^4+\nonumber\\
&&(\partial_{\mu}\sigma)^{\dagger}(\partial^{\mu}\sigma)-m_{\sigma}^2|\sigma|^2+\frac{\beta}{4}|\sigma|^4+\frac{\delta}{4}|H|^2|\sigma|^2,
\end{eqnarray}
where $\alpha$, $\beta$ and $\delta$ are constant parameters. The parameter
$\delta$ is related with the mixing between the electroweak and the $B-L$ groups
and we are going to work in the limit where $\delta \ll\beta$. This is the most
simple realization of $U(1)_{B-L}$, where the electroweak gauge group is not
mixed with the $B-L$ one, thus, the results presented in here would be
considered as a first approximation of the full $SU(2)_L\times U(1)_Y\times
U(1)_{B-L}$ gauge theory.

We still have not identified the breaking scale of $U(1)_{B-L}$ and, in this
context, we will not be able to determine it, unless an experimental evidence of
a $B-L$ gauge boson appears. Hence, we have two possibilities for the breaking
scale: {\it i)} if $\langle H \rangle \ll \langle \sigma \rangle$ we end up in
the class of models where $m_{Z_{B-L}}\gg m_Z$ and the physics related to these
gauge groups is decoupled; and {\it ii)} if $\langle H \rangle \sim \langle
\sigma \rangle$, the physics involved in this scenario is such that the gauge
bosons are not decoupled. Moreover, the non evidence of new physics hits also to
this class of models, meaning that the probable effects might be suppressed in
some manner. In this work, we consider that $m_{Z_{B-L}}\lesssim m_Z$, the
motivation of this proposal is based on the analysis made by Ref.~\cite{Gor08,
Boy08}, where it was pointed out that a RHN mass of the order of keV is a good
candidate to be the DM particle, and, in this context, our above assumption
might lead us to this DM scheme.

By imposing the spontaneous breaking of the local $U(1)_{B-L}$ symmetry, a
massive neutral gauge boson, $Z_{B-L}$,  emerges with a mass 
\begin{equation}
M^2=8g^2_{B-L} \langle \sigma \rangle ^2,
\end{equation}
where $g_{B-L}$ is the coupling constant associated to $U(1)_{B-L}$ and
$\langle\sigma\rangle$ is the vacuum expectation value of the $\sigma$ field. 
In addition, $Z_{B-L}$ couples to matter through the
coupling $-g_{B-L} Z_{\mu}^{B-L} J^{\mu}_{B-L}$, where the vector-like $B-L$
current is given as,
\begin{eqnarray}\label{current}
J^{\mu}_{B-L}=\bar{\nu}_L \gamma^{\mu}\nu_L + \bar{e}_L \gamma^{\mu} e_L +
\bar{e}_R \gamma^{\mu} e_R + \bar{N} \gamma^{\mu} N~.
\end{eqnarray}

With this simple structure, we can implement several searches of $B-L$ currents
in different levels. First, a realistic model for RHN as dark matter must
satisfy the cosmological limits which are set due to several observations. So
next, we are going to review the applicable cosmological constraints for this
model.

\section{Cosmological Dark Matter Constraints}

Let us start the discussion of the model by looking at the cosmological
constraints. As it is pointed out in Ref.~\cite{Bez10},  models with RHN as
DM must satisfy the constrains given mainly by: \textit{a)}  the structure
formation requirement, \textit{b)} X/$\gamma$-ray observations and, \textit{c)}
the abundance of these particles at present time. 

If the mass of the DM candidate is too light, the observed structure of the
Universe would be eliminated by a too hot DM. The most restrictive bound on the
mass of the DM candidate is coming from the Lyman-$\alpha$ forest, where the
bound constraints the velocity distribution of the DM particles from the effect
of their free streaming on the formation of the structure on Lyman-$\alpha$
scales. The bound on our lightest RHN mass is thus given by~\cite{Bez10},
\begin{eqnarray}
M_1 > 1.6 \mbox{ keV}~.
\end{eqnarray}
Therefore, we are going to work with this value, not to spoil the large
structure formation.

\begin{figure}[t!]
\centering
\includegraphics[scale=0.45]{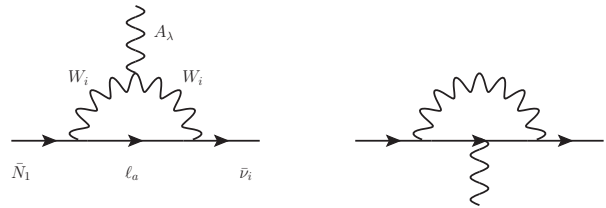}
\caption{Feynman diagrams which contribute to $N\rightarrow \nu\gamma$ decay.}
\label{fig1}
\end{figure}

On the other hand, a general feature of this model is that the RHN, the Dark
Matter candidate, does decay into SM particles. This characteristic appears
basically due to the fact that RHN could transform into an active one via the
mixing parameter,
\begin{eqnarray}
\theta_I =\sum_{\alpha=e,\mu,\tau}\frac{\mathcal{K}_{\alpha I}v^2}{M_1},
\end{eqnarray}
where $v$ is the SM Higgs vacuum expectation value. Thus, one can compute the
decay rate of RHN into charged leptons plus a SM neutrino ($\ell^+\ell^- \nu$)
or into three SM neutrinos in order to check a natural DM requirement: the
lightest RHN should be long lived and, furthermore, its lifetime has to be
longer than the age of the universe. Nevertheless, this constraint is less
dominant compared to what it results from the X/$\gamma-$rays observation. The
radiative decay $N\rightarrow \nu\gamma$, induced at one loop level (FIG.~1),
produces a narrow line in the X-ray spectrum of astrophysical
objects~\cite{Dol02, Aba01b}. The width of this decay is given by~\cite{Boy09},
\begin{eqnarray}
\Gamma_{N\rightarrow \nu\gamma}\simeq \frac{9 \alpha G_F^2}{1024 \pi^4}\sin^2(2\theta_1)M_1^5,
\end{eqnarray}
where $\alpha$ is the fine-structure constant, and $G_F=1.166\times 10^{-5}$
GeV$^{-2}$ is the Fermi constant. This width can restrict the mixing parameter
when we consider the upper bound found in Ref.~\cite{Bez10}, which reads as,
\begin{eqnarray}
\Gamma\lesssim 9.9\times 10^{-27}\mbox{ sec}^{-1},
\end{eqnarray}
and it is translated to
\begin{eqnarray}
\theta_1^2\lesssim 1.7 \times 10^{-6} \left(\frac{1.6 \mbox{ keV}}{M_1} \right)^5.
\end{eqnarray}
Within the context of the parameters of the model, this result can be seen as a
bound to the Yukawa parameters. Therefore, no restriction on the $B-L$ coupling
and the mass of the associated gauge boson can be made so far.

Finally, the abundance of RHN at the present time can be calculated by using~\cite{Bez10},
\begin{eqnarray}
\frac{\Omega_N}{\Omega_{DM}} = \frac{135 \zeta(3) }{4\pi^4g_{*f}}  \frac{M_1}{S}\frac{ s_0}{ \Omega_{DM} \rho_c}
\end{eqnarray}
where $\Omega_{DM}=0.105 \,\,h^{-2}$, $s_0 = 2889.2$ cm$^{-3}$, $\rho_c = 1.05368\times 10^{-5} \,\,h^2$ GeV cm$^{-3}$, $S$ the entropy release and $g_{*f}$ is the effective number of degrees of freedom immediately after freeze-out . Therefore, for a 1.6 keV RHN mass, we have,
\begin{eqnarray}
\frac{\Omega_N}{\Omega_{DM}} \simeq \frac{1 }{S} \left( \frac{10.75}{g_{*f}} \right)\left(\frac{M_1}{ 1.6 \mbox{ keV}}\right) \times 160,
\end{eqnarray}
where we have chosen $g_{*f}=10.75$, this value corresponds to the case where there are only SM particles and the freeze-out have happened below 100 MeV. In order to not overclose the universe we need a dilution factor of the order of,
\begin{eqnarray}\label{Sneeded}
S \simeq  160 \times \left( \frac{10.75}{g_{*f}} \right)\left(\frac{M_1}{ 1.6 \mbox{ keV}}\right),
\end{eqnarray}
we can achieve this requirement by considering that heavier RHNs could generate
such amount of entropy. Let us consider that only the heaviest right handed
neutrino, $N_3$, is the author of such entropy. In the case where the entropy
generation is large, the calculation gives~\cite{Bez10} ,
\begin{eqnarray}\label{Seq}
S \simeq 0.76 \times \frac{g_N}{2} \frac{\bar g_*^{1/4} m_H}{g_* \sqrt{\Gamma M_{Pl}}},
\end{eqnarray}
where $g_N=2$ corresponds to the number of degrees of freedom of $N_3$, $\bar
g_*$ and $g_*$  are the properly averaged effective number of degrees of freedom
during $N_3$ decay and at freeze-out, and $m_H$ is the heavy neutrino mass.
Therefore, Eq. \eqref{Sneeded} and \eqref{Seq} can be solved in order to
constraint $m_H$, nevertheless, the entropy generation should end before the big
bang nucleosynthesis occurs (BBN), saying, the reheating temperature should
satisfy the limits~\cite{Asa06, Kaw00, Han04} $0.7 < T_R < 4$ MeV. This
temperature is approximatively~\cite{Sch85},
\begin{eqnarray}\label{temperature}
T_R \simeq \frac{1}{2} \left( \frac{2 \pi^2 \bar g_*}{45} \right)^{-1/4}\sqrt{\Gamma M_{Pl}}.
\end{eqnarray}
Combining Eq. \eqref{Sneeded}, \eqref{Seq} and \eqref{temperature}, we find a constraint on $m_H$ to be,
\begin{eqnarray}\label{mheavy}
m_H > \left(\frac{M_1}{1.6 \mbox{ keV}} \right) (2.5 \div 15) \mbox{ GeV}
\end{eqnarray}

Although, it has been pointed out by some authors~\cite{Bez10}, that once we
have a constraint on the mass of the heavy RHN, it is possible to get a bound on
the mass of the extra gauge boson. In order to do so, one needs to make use of
the following fact: whether a heavy RHN is decoupled, the decoupling must
happens when the freeze-out temperature is higher than the mass of this heavy
RHN.
So, we just need to estimate the freeze-out temperature by considering that the mean free path of RHN should be equal to the Hubble scale.
In order to estimate the moment when RHN decouples to the SM particles, we can
use that the processes involved 
($\bar N_1 N_1 \leftrightarrow e^+e^-$, etc.) which are analogous to those for the
usual SM neutrinos; the only difference appears in the coupling and in the mass
of the $B-L$ gauge boson, therefore,
\begin{eqnarray}
\sigma_{N_1\bar N_1}\approx \sigma_{\nu\bar\nu} 
\left( \frac{g_{B-L}}{g}\frac{M_W}{M}\right)^4
\end{eqnarray}
where $g$ is the $SU(2)_L$ gauge coupling, and $M_W$ is the mass of the $W$
boson. And so, the freeze-out temperature, $T_f$, would be
\begin{eqnarray}
T_f\sim g_*^{1/6}\left( \frac{g}{g_{B-L}}\frac{M}{M_W}\right)^{4/3} (1 \div 2) \mbox{ MeV}.
\end{eqnarray}
Hence, one gets a bound for  $B-L$ gauge boson mass which reads as,
\begin{eqnarray}
M> \left(\frac{g_{B-L}}{g}\right)(6 \div 10) \mbox{ TeV},
\end{eqnarray}
which is remarkably close to the limit set by collider searches by a few
percentage. Furthermore, this result matches in the limit when the gauge bosons
couple equally, to the result presented in Ref.~\cite{Bez10}. Nevertheless, 
we have no idea about the absolute scale of $g_{B-L}$ so far. This is important 
since it defines whether $B-L$ physics could be relevant at any given scale.
For instance, for a sufficiently small value of the coupling, say $g_{B-L}\sim
\mathcal{O}(10^{-3}$), one could resize the lower bounds on $M$ down to few GeV scale,
suggesting that former experiments should have be sensible to the involved
physics. 
It is, therefore,  of a clear
interesting to know how big the value of $g_{B-L}$ is, in order to know whether
its effects can be seen in future experiments.

\section{Constraining the model using PETRA Experiment}

Past low energy experiments can help us to upper bounding the value of
$g_{B-L}$. A very clean channel where an extra neutral gauge boson can appear is
in $e^+e^-$ scattering, and this reaction was studied by the PETRA
experiment. We now reanalyze PETRA results in order to explore for a possible
low scale mass of $Z_{B-L}$. This limit would give us a hint about the
values of $g_{B-L}$ or $\langle \sigma \rangle$, and therefore we can stretch
the limit in order to fulfill the DM constraints. Although, LEP and Tevatron
experiments, and eventually LHC data,  can also be included in the analysis, the
known bound already
stated in Eq.~(\ref{zpbound}) get particularly soften for small values of the
$B-L$ coupling, as we mentioned above. Thus, we would expect little improvement
of the combined analysis,  whereas it would become more involved. For
simplicity, we left this analysis for a further work. In the following, we will
use this experiment to restrict whether the mass of the gauge boson or the gauge
coupling.

PETRA results had been well explained theoretically by using the simple Bhabha
scattering process. The existence of a light and weakly coupled
$Z_{\mu}^{B-L}$ can be strongly restricted
by checking the contribution of whatever extra neutral gauge boson to the Bhabha
cross section. So, we perform the computation of $e^-e^+ \rightarrow e^-e^+$
scattering to establish a bound on $g_{B-L}$ and $M$. The relevant Feynman
diagrams for the process are pictured in FIG.~2.

\begin{figure}[t!]
\centering
\includegraphics[scale=0.3]{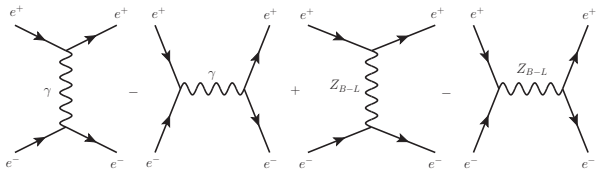}
\caption{Feynman diagrams for  $e^-e^+ \rightarrow e^-e^+$ 
in the $SU(2)_L\times U(1)_Y\times U(1)_{B-L}$ theory.}
\label{fig:}
\end{figure}

Thus, in the limit $E\gg m$, Bhabha's expression receives a contribution from
$Z_{B-L}$ and the result can be read as, 
\begin{widetext}
\begin{eqnarray}\label{bhabha corregido}
\frac{d\bar\sigma}{d\Omega}=
\left( \frac{d\bar\sigma}{d\Omega} \right)_{B} \left[ 
1 + \frac{g_{B-L}^2}{e^2}~\Xi\left(\frac{M^2}{E^2},\cos\theta\right) +
\mathcal{O} \left( \frac{g_{B-L}^4}{e^4}\right) \right],
\end{eqnarray}
where
$\left( \frac{d\bar\sigma}{d\Omega} \right)_{B}$ is the Bhabha's differential
cross section and the correction due to the extra gauge boson is enclosed in
the $\Xi$ function, such that,  
\begin{eqnarray}
\Xi\left(x,y\right) = 
\frac{-16(1-y)(-216+4x(9+3y-y^2-3y^3)+8y^2(-14+8y+y^2))
}{
(x-4)(x+2-2y)(-216+8y^2(-14+8y+y^2))}.
\end{eqnarray}

\begin{figure*}[ht]
\includegraphics[scale=0.44]{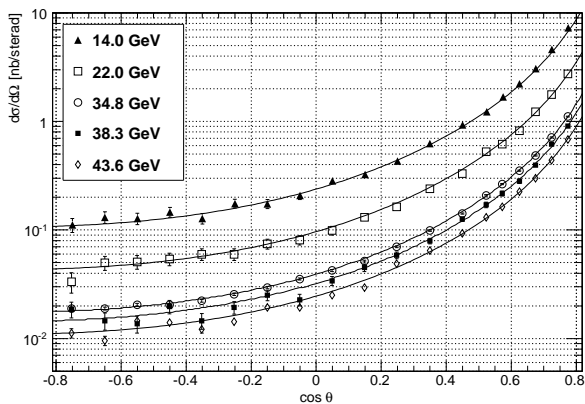}
\includegraphics[scale=0.45]{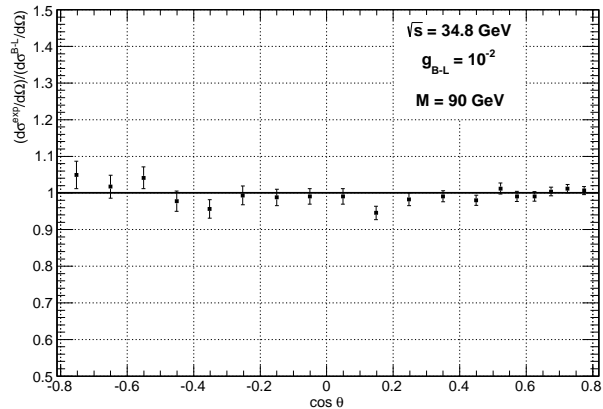}
\caption{Correction to PETRA experiments. On the left panel we plotted the results of the $SU(2)_L\times U(1)_Y \times U(1)_{B-L}$ with the experimental data. On the right panel we plotted the ratio of the experimental measure and the theory described here, for $\sqrt{s}= 34.8$ GeV, due to the experimental data is more accurate at this center of mass energy.}
\end{figure*}
\end{widetext}

A numerical analysis of~\eqref{bhabha corregido}  shows that for any value
$g_{B-L} \lesssim 10^{-2}$ ~\cite{App03,Thesis},  PETRA results can be
satisfied independently of the value for $M$, provided $M\gg E$. Indeed,
by using  $g_{B-L}= 10^{-2}$, for instance, it is possible to reproduce the
PETRA results~\cite{PETRA}, as it is shown in the FIG.~3, for $M=$ 90 GeV.

For such an small value for $g_{B-L}$, LEP limit becomes less restrictive, meaning that we can allow masses of the $Z'$ as low as 90 GeV and remain still consistent with the previous observations. If we consider, naively, that $M = 90$ GeV, we are also considering that the breaking scale of $U(1)_{B-L}$ is occurring at 
\begin{eqnarray}
\langle \sigma \rangle \sim 3 \mbox{ TeV,}
\end{eqnarray}
which is a reachable scale for the LHC.

On the other hand, the mass of the right handed neutrino is not fully controlled by $\langle \sigma \rangle$. Considering the Majorana term in Eq.~\eqref{Yukawa},
we can still have a RHN mass of the order of keV by choosing the Yukawa
parameter to be $ y_1 \sim 2.5\times 10^{-10}$. For the heaviest RHN, a
Yukawa larger than $10^{-3}$ would be enough to satisfy the bound on
Eq.~(\ref{mheavy}).
Certainly,  on this path, one gives up
naturalness on the see-saw mechanism, but gain the possibility of accounting for
DM.

\section{Concluding remarks}

In this paper we have analyzed a simple extension of the electroweak gauge group
by adding an extra $U(1)_{B-L}$. This combination of barion and lepton numbers
makes the model anomaly free. Moreover, with this combination of quantum
numbers, we have been allowed to write the RHN field and, by choosing
$Y_{B-L}=2$ for an extra Higgs field, $\sigma$, we have been able to write a
Majorana mass term for neutrinos when this extra Higgs field acquires a vacuum
expectation value. Spontaneous breaking of $U(1)_{B-L}$ gauge leaves a
a massive $Z_{B-L}$ gauge boson, which related to the
breaking scale, and the coupling constant associated to $U(1)_{B-L}$.
By adding the corresponding  RHN fields, we can address the problem of the DM
content in the universe according to Ref.~\cite{Gor08, Boy08}. Thus, we studied the
minimum conditions to recover the conclusions done for $\nu$MSM in the DM point
of view. In order to say something about neutrinos as dark matter, we have
analyzed the cosmological constraints. These constraints had required to the
lightest RHN to have a mass of about few keV in order to
satisfy the observed structure formation. The X/$\gamma-$ray observation has set
a bound on the mixing between active and sterile neutrinos to be
$\theta_1^2\lesssim 10^{-6}$. In order to produce the correct abundance of RHN
as DM, we have restricted the mass of the heavy RHN to be of the order of
few GeV. The imperative decoupling of the heavy RHN from the
lightest one has set a limit on the mass of the $Z_{B-L}$, which is comparable
to that obtained from LEP data; nevertheless, the relation found is not
conclusive since we still have the possibility of a small $B-L$ gauge coupling,
such that, we could still have a consistent light $Z_{B-L}$ gauge boson. 
Thus, in order to find an upper bound to $g_{B-L}$ we analyzed the possibility
that the net effect of $U(1)_{B-L}$ could be suppressed in the process $e^+e^-
\rightarrow e^+e^-$; in particular, since it must appears as a contribution into
the Bhabha scattering. The simplest experiment which we can compare to is PETRA
which has already measured exhaustively this scattering process. A combined 
analysis for LEP, Tevatron and eventually LHC experiments is left for
a future work. The result indicates that, if the  coupling constant associated
to $U(1)_{B-L}$ is of the order of $10^{-2}$ or less, the contribution given by
$Z_{B-L}$ is indeed subleading to the level of being mostly negligible. With this
condition, we are also imposing that the breaking of $U(1)_{B-L}$ should occurs
at about  3 TeV or higher, a sizable value that come into the energy region of
interest for the LHC.

\begin{acknowledgments}
The authors thanks the referee for helpful comments. This work was supported
partially by CONACyT, M\'exico, grants 54576 and 132061.
\end{acknowledgments}

\end{document}